\documentstyle{mn}

\begin{document}

\def\sameauthor{\underbar{\qquad\qquad}.}
\def\msun{{\rm M}_\odot}
\def\rsun{{\rm R}_\odot}
\def\halpha{H$\alpha$}
\def\hbeta{H$\beta$}
\def\hgamma{$H\gamma$}
\def\by{$\times$}
\def\av#1{\langle#1\rangle}
\def\gsimeq{\, \lower 0.5ex\hbox{$\buildrel {\scriptstyle>} \over {\scriptstyle\sim}$}\, }    
\def\lsimeq{\, \lower 0.5ex\hbox{$\buildrel {\scriptstyle<} \over {\scriptstyle\sim}$}\, }    
\def\gradi{\ifmmode{^\circ}\else$^\circ$\fi}
\def\reference{\parskip 0pt\par\noindent\hangindent 0.5 truecm}

\title{Studies of magnetic and suspected-magnetic southern white dwarfs}
\author[Gary D. Schmidt, St\'ephane Vennes, D.T.~Wickramasinghe, \& L. Ferrario]
     {Gary~D.~Schmidt$^{1,2}$, Stephane Vennes$^1$, D.T.~Wickramasinghe$^1$, \& L. Ferrario$^1$ \\
     $^1$Department of Mathematics, Australian National University,
          Canberra\\
     $^2$Steward Observatory, The University of Arizona,
        Tucson, Arizona}

\date{Accepted. Received}
\pagerange{\pageref{firstpage}--\pageref{lastpage}}
\pubyear{2001}

\maketitle

\begin{abstract}

Optical spectrophotometry and circular spectropolarimetry are presented for
several candidate magnetic white dwarfs that were identified during the
Hamburg/ESO survey for bright QSOs. Two objects, HE~1211$-$1707 and
HE~1043$-$0502, are shown to be rare examples of white dwarfs that show
neutral helium lines in a high magnetic field, in these cases $\sim$50~MG and
$\sim$800~MG, respectively. The former is also found to be rotating with a
period of $\sim$2~hr. HE~1045$-$0908 is a hydrogen-line star with a polar field
strength of $\sim$20~MG, spinning with a period in the range $\sim$$2-4$~hr.
Attempts at modeling the limited amount of phase-resolved data that is
available suggest that the field structure on this star departs substantially
from a simple centered dipolar geometry. Two rather cool white dwarfs with
unidentified broad absorption spectral features, HE~0236$-$2656 and
HE~0330$-$0002, are confirmed polarimetrically to be magnetic.  Line
identifications for these stars are not yet possible, but the atmospheres are
probably helium-rich, with spectral features formed by trace compounds of
hydrogen, carbon, and perhaps other metals.  HE~0003$-$5701 and HE~0338$-$3853
were proposed by Reimers et al. (1996) alongside two similar objects to be
magnetic DB white dwarfs, with Zeeman-split lines of helium in magnetic fields
all near 20~MG. However, our observations show that the first two, and by
extension all four, are nonmagnetic white dwarf + cool dwarf pairs. They are
deserving of study in their own right as possible close binaries.  Finally, a
lack of circular polarisation suggests that HE~0000$-$3430 and HE~0127$-$3110
are also nonmagnetic. HE~0000$-$3430 appears to be a featureless DC white
dwarf over the spectral range observed here, while the sole absorption line
near 5890~\AA\ in HE~0127$-$3110 could be either He~I $\lambda$5876 or the
Na~I D doublet, but there are difficulties with either interpretation.

\end{abstract}


\section{Introduction}

Magnetic white dwarfs offer an important glimpse into the roles that magnetism
may play in the formation and evolution of stars with moderate mass. At the
same time, they provide unique laboratories for studying the behavior of matter
in fields far stronger than can be obtained terrestrially. More than 5 dozen
examples have been found with field strengths between $\sim$$3\times10^4$~G and
$10^9$~G, and the list of atomic and molecular species represented includes
virtually every substance seen among white dwarfs in general, including H, He,
Na, Mg, Ca, C$_2$, and additional molecules that are as yet unidentified.

Modeling of spectroscopic and spectropolarimetric data on magnetic stars is a
powerful technique for gaining information into the field distributions over
the stellar surfaces, particularly for the objects in which the observations
can be phase-resolved over a rotational cycle. This information is crucial for
evaluating alternatives for the origin and evolution of the fields, and for
relating magnetic structures found among one class of star to other stages of
evolution. Unfortunately, a lack of thorough observational material has
traditionally handicapped studies of white dwarfs in the southern hemisphere.
During the course of a spectropolarimetric survey for magnetic fields among
southern white dwarfs, special attention was paid to several known or proposed
magnetic examples.  This paper reports the results on 9 objects which emerged
from the Hamburg/ESO survey for bright QSOs (e.g., Wisotzki et al. 1995).

\section{Observations}

The data presented here were acquired with the Steward Observatory CCD
Spectropolarimeter (Schmidt, Stockman, \& Smith 1992) attached to the 74-inch
reflector on Mt. Stromlo.  Since that description was published, the
instrumental performance has been upgraded with an improved camera lens and a
thinned, back-illuminated LORAL 1200\by800 CCD with near-unity quantum
efficiency and $<$6$e^-$ read noise.  For the current application, the $\it
f$/18 Cassegrain telescope beam was adapted to the $\it f$/9 spectrograph
optics with a small converging lens placed ahead of the slit. The instrument
was configured for circular spectropolarimetry over the region
$\lambda\lambda$$4220-7300$~\AA\ with $\sim$9~\AA\ resolution, and data were
acquired in multiple waveplate sequences of typical duration $12-14$ minutes.
Shorter sequences were used for HE~1211$-$1707 in order to resolve variations
over the known short rotation period (Reimers et al. 1996).  Because it was not
possible to align the instrument to the parallactic angle and the arrival of
clouds occasionally prevented obtaining a nightly flux calibration standard,
the large-scale absolute flux calibrations of the program stars are probably
accurate to only $\sim$20\%. This difficulty has no systematic effect on the
polarimetry, since both senses of polarisation for the determination of a
Stokes parameter are accumulated simultaneously in parallel spectra on the CCD.
Terrestrial absorption features were removed using spectra of hot white dwarfs
taken with the same instrumental setup, but because of differences in airmass,
cancellation is not always complete beyond 7000~\AA.  A summary of the
observations is presented as Table~1.  Included is a measure of the circular
polarisation summed over the entire spectrum and the monochromatic Palomar AB
magnitude $m_{5500}$ determined from the slit spectrum.

\section{Results}

\subsection{Magnetized neutral helium at 50~MG and 800~MG in
HE~1211$-$1707 \& HE~1043$-$0502}

HE~1211$-$1707 presents an interesting spectrum with a series of time-variable
absorption features that are clearly indicative of rotation (Reimers et al.
1996). Our data sets, obtained on two nights separated by a few days, each
consist of a sequence of 15 consecutive 6 min observations spanning $\sim$100
min.  The circular polarisation summed over the spectrum and over all
rotational phases is not strong but significant at $v\sim1\%$ (Table~1),
confirming the presence of a magnetic field.  Figure~1 presents the spectral
flux series from 25 Feb., shown per unit frequency to highlight the structure.
Because of the comparative faintness of the star, the observations were
averaged by pairs prior to plotting, with UT progressing upward as noted at
the right. The initial two spectra reveal the broad features near
$\lambda\lambda$4700, 5250, 5750 discussed by Reimers et al., with maximum
depths of $\sim$20\%. By 30~min into the sequence, all but the $\lambda$5750
feature have disappeared, but that line has sharpened significantly.  The next
two spectra, at $45-55$ min into the series, reveal the emergence of shallow
depression centered near $\lambda$5050. The last (single) exposure shows a
return of the structure seen at the beginning of the sequence, and suggests
that nearly a full spin cycle has been completed.  We infer that the rotation
period is $P\sim2$~hr. The sequence from 22 Feb. mimics Figure~1, offset in
phase. The spectrum-summed circular polarisation from the individual
observations shows a roughly sinusoidal modulation from $v\sim0\%$ to +3\%,
with a period of $\sim$$100-120$~min and a peak near the maximum strength of
the $\lambda$5050 feature. Unfortunately, the lack of coverage of a complete
cycle in either series or in the data of Reimers et al. precludes phasing the
two nights, so the spin period cannot be quoted more accurately.

The $4300-7300$~\AA\ flux distribution  of HE~1211$-$1707 implies a surface
temperature of $T_{\rm eff}\sim12,000$~K, considerably cooler than that
indicated by the {\it IUE\/}/optical flux ratio (Reimers et al. 1996) and
suggesting line blanketing of the optical continuum.  The temperature regime
is one where either a hydrogen- or helium-atmosphere white dwarf might be
expected. Reimers et al. (1996) noted some evidence for Balmer lines in a
magnetic field of $\sim$80~MG, but the features could not be identified
unambiguously and other features remained unidentified altogether. Our mean
spectrum, shown in Figure~2, depicts a large amount of structure, including
the aforementioned broad features and a very diffuse dip extending from
$\sim$6200~\AA\ to the red.  The rather modest $0-3\%$ range for the
phase-modulated circular polarisation, together with the fact that $v$ does
not exceed $\sim$1.5\% in the coadded polarisation spectrum, suggests that the
field strength is not extremely high.  There is some evidence for a slight
decline in circular polarisation with wavelength.

Very recently, calculations of many of the transitions of neutral helium have
become available for a wide range in the applied magnetic field. In Figure~3
we reproduce the behavior of the principal lines for $0<B<900$~MG, taken from
the calculations of Becken \& Schmelcher (1998, 2000, 2001) and Becken,
Schmelcher, \& Diakonos (1999).  We see here that at moderate field strengths,
$B\lsimeq50$~MG, the range $5000-7000$~\AA\ is dominated by the Zeeman triplet
of 2$^3$P~$-$~3$^3$D $\lambda$5876, rapidly-moving $\sigma^-$ components of
2$^1$P~$-$~3$^1$D $\lambda$6678, and $\sigma^+$ components of
2$^1$S~$-$~3$^1$P $\lambda$5015. Due to the quadratic Zeeman effect, the $\pi$
components of $\lambda$5876 are shifted to $\sim$5750~\AA. The $\sigma^\pm$
features reach wavelengths of $\sim$6500~\AA\ and $\sim$5100~\AA,
respectively, and each of these will be smeared over a few hundred angstroms
even in the $\times$2 pole-to-equator field spread of a centered dipole. Thus,
the $\lambda$5876 triplet offers an attractive match to the three
long-wavelength features at $\sim$5250, 5750, and 6200~\AA\ in our summed
spectrum of HE~1211$-$1707, for a field strength of $B\sim50$~MG.  We note
that Jordan (2001) has also recently advanced a helium explanation for the
features of this star.

We have investigated this interpretation through spectral modeling using a
transfer code for polarised radiation. The approach follows that described by
Wickramasinghe \& Martin (1979) but uses an atmospheric structure appropriate
for a helium-atmosphere white dwarf. Unfortunately, the new calculations of
He~I transitions mentioned above do not provide oscillator strengths and have
not yet been completed for all transitions of interest. Therefore, our current
models are based on extrapolations of the perturbation calculations of Kemic
(1974), which characterize 18 transitions up to a limiting field strength of
10~MG or 20~MG, depending on transition. Wherever possible, the $B(\lambda)$
curves of Kemic were extended to higher fields using the newer calculations
(i.e., for the triplets $\lambda\lambda$5876, 4714, 4472, and 4026; plus the
singlets $\lambda\lambda$4923, 4388, 4144, 5017, 3965, and 3972).  Becken and
collaborators also include some transitions between singlet and triplet states
with positive and negative $z$ parity for $\Delta M=0$, $\pm1$, and $\pm2$.
Nevertheless, some transitions have not yet been updated.  For the wavelengths
of these, linear extrapolations were applied beyond Kemic's (1974) limiting
field strengths. Oscillator strengths are all from Kemic, except when the
field strength exceeded his limit; then the value at the limiting field
strength was used.  Of course, attempting to construct models with such a
patchwork database is very unsatisfactory, and mismatches between the
observations and calculations are inevitable, particularly for the higher
field surface areas. Model polarisation spectra would be especially
unreliable.  We intend to readdress this question when more complete atomic
calculations become available.

Two models which are in reasonable agreement with the mean observed spectrum
are shown in Figure~2, labeled according to the dipolar field strength $B_d$,
the assumed viewing inclination and azimuth with respect to the field pattern,
and the longitudinal displacement of the dipole, in units of the stellar
radius.  The centered model uses $B_d=49$~MG and the offset model has a
visible pole with $B=57$~MG, so the maximum field strength on the stellar
surface is near 50~MG for both models, as expected from the $B(\lambda)$
curves.  The apparent detection of Ly$\alpha$ $\sigma^+$ in the {\it IUE\/}
spectrum of HE~1211$-$1707 (Reimers et al 1996) suggests that the atmosphere
may also contain small amounts of hydrogen. The field strength quoted for that
feature is $\sim$80~MG, but Ly$\alpha$ is extremely insensitive to field, and
we find the agreement between the models and observed optical spectrum in
Figure~2 convincing evidence that HE~1211$-$1707 presents features of neutral
helium in a magnetic field that ranges to little more than 50~MG over the
stellar surface.

Several very broad, unidentified features in the spectrum of HE~1043$-$0502
led Reimers et al. (1998) to suggest that the star was magnetic.  The
substantial broadband circular polarisation measured on each night of our
observations, $\av{v}\sim+1.5\%$, confirms this conclusion, and values as high
as $v=4$\% are observed in certain spectral regions of the data shown in
Figure~4. Because of the faintness of the object, spectrum and/or polarisation
time variations, if they are present, are too subtle to search for stellar
rotation, so the combined results from both nights are shown.  Reimers et al.
(1998) pointed out that the shape of the deep $\lambda$4450 feature resembles
the asymmetric absorption lines in GD~229, which have been shown to be due to
a large number of neutral helium transitions reaching wavelength minima in
magnetic fields $300 \lsimeq B \lsimeq 700$~MG (Jordan et al. 1998).  An
additional very broad and shallow depression appears from
$\sim$$5000-6000$~\AA\ in our spectrum of HE~1043$-$0502 and a weak, narrower
feature exists in the data of Reimers et al. around 4000~\AA. The continuum
shape suggests a temperature ($\sim$15,000~K) that is appropriate for the
existence of He~I.  However, Reimers et al. were unable to achieve a
satisfactory fit to magnetic helium for the transitions which had been
computed at that time.

From a comparison of the spectrum of HE~1043$-$0502 with the $B(\lambda)$
curves in Figure~3, it can be seen that the deep $\lambda$4450 feature lies
only slightly longward of the $\lambda4291$ short-wavelength turnaround of He~I
$\lambda$5015 $\pi$ (2$^1$S$_0-3^1$P$_0$) at a field strength of 359~MG. If
we interpret the tremendous width of the observed profile as being due to
magnetic broadening, the field strength must range up to $\sim$800~MG and down
to $\sim$450~MG over the stellar surface. In this regime, the $\sigma^+$
component of the same zero-field line (2$^1$S$_0-3^1$P$_{-1}$) is moving
quickly to the red from its minimum near $\lambda$4812 (251~MG), and is a
likely explanation for the shallow depression between 5000~\AA\ and 6000~\AA.
The only other feature expected is an extremely smeared trough covering
$\sim$$6300-7000$~\AA\ arising from the $\sigma^+$ component of $\lambda$5876
(2$^3$P$_0-3^3$D$_{-1}$) as it moves back from a wavelength maximum near
$\lambda$7143 (258~MG).  This may also be present, weakly, in the observed
flux spectrum in Figure~4. Unfortunately, the limitations that plagued our
modeling of HE~1211$-$1707 -- the lack of oscillator strengths and
incompleteness of wavelengths for some lines -- are fatal at these much higher
field strengths, and we must be satisfied for the time being with visual
matches to the $B(\lambda$) curves.  However, we find these assignments
convincing and conclude that HE~1043$-$0502 has a field of $\sim$$450-800$~MG
over the surface, higher even than GD~229, and the strongest field yet
measured on a DB white dwarf.

\subsection{The rotating magnetic DA HE~1045$-$0908}

A rich Zeeman spectrum of hydrogen was detected in HE~1045$-$0908 by Reimers
et al. (1994) and modeled as arising from a dipolar field distribution of
strength $B_d\sim31$~MG viewed nearly equator-on.  The exposure time of
Reimers et al. was unspecified, but it is clear in our data series shown in
Figure~5 that systematic variations occur in both spectral flux and circular
polarisation over the 1~hr duration of the observations.  At the beginning of
the sequence, the lines are ill-defined and circular polarisation is modest,
reaching at most $v=\pm5\%$ in the $\sigma^\mp$ components of \hbeta. The
features become even more diffuse in the next spectrum, but sharpen and
increase in polarisation thereafter.  By the last two exposures, even the
$\sigma^-$ components of \halpha\ are obvious in total flux, and polarisation
excursions reach nearly $\pm$10\%.  These characteristics can be understood
intuitively in terms of the rotation of an oblique magnetic field pattern,
with the initial observations of the sequence being dominated by a wide spread
in field strength, while the better-defined features in later measurements
depict a more restricted range.  Depending on the geometry, Figure~5 therefore
probably represents about one-quarter to one-half of a full rotational cycle,
implying that the spin period is likely in the range $\sim$$2-4$ hr.

The mean observed spectrum of HE~1045$-$0908 is compared to a series of
spectral models in the top panel of Figure~5. All of the models are seen at
low inclination, $20\gradi < i < 30\gradi$ and at a small azimuth to the
dipole axis.  This basic geometry is indicated by the general lack of
weak-field Zeeman features in the observed spectra.  The models differ in the
longitudinal offset of the assumed dipole field structure and in the
corresponding polar field strength, which range from from $B_d=19$~MG; $\Delta
R/R=0.0$ to $B_d=9$~MG; $\Delta R/R=0.3$.  Of the 3, the highly offset geometry
best reproduces the width and shape of \halpha\ $\pi$ and $\sigma^+$, but
predicts a too localized and displaced $\sigma^-$ component.  Differences
around \hbeta\ are less pronounced.

Difficulties appear when these particular models are applied to the spectrum
and polarisation variations. Here we see that the observed \halpha\ $\sigma$
components broaden out almost completely at some phases. This behaviour is
inconsistent with spectral variations expected from centered or longitudinal
($z$)-offset dipole models. By $z$-offsetting the dipole, the field spread
increases as viewing angle increases, and the contributions from lower field
regions tend to dominate the spectrum. Note also that in these models the
field direction changes very quickly over the surface so that as the viewing
angle increases, the sign of circular polarisation of the $\sigma$ components
also changes. This is not observed.

We conclude that, without high-quality observations over a complete rotation
cycle, attempts at detailed modeling of the field geometry over the stellar
surface of HE~1045$-$0908 are probably premature.  We suspect that more general
offsets (transverse as well as along the dipole axis) or substantial
departures from a dipolar geometry may be required to explain the spectral and
polarisation variations seen in the data.  This star may be a particularly
interesting target for modeling using modern genetic algorithms (e.g., Hakala
1995) for approaching global optimization of a large number of parameters that
describe a magnetic field structure.  We note that several efforts at modeling
phase-dependent variations of several other magnetic white dwarfs point toward
substantial departures from simple magnetic geometries, with quadrupoles,
offset dipoles, and even magnetic ``spots'' being demanded by the data (e.g.,
Schmidt et al. 1986; Maxted et al. 2000).

\subsection{HE~0236$-$2656 \& HE~0330$-$0002: Cool magnetic white dwarfs with unidentified features}

Together with HE~1043$-$0502 discussed in $\S$3.1, HE~0236$-$2656 and
HE~0330$-$0002 were flagged as suspicious by Reimers et al. (1998) on the
basis of broad, unidentified absorption features in their spectra.  As
indicated in Table~1, both stars show substantial broadband circular
polarisation and thus can now be confirmed as magnetic white dwarfs.  Each
star was observed twice separated by a few days, and in each case consistent
results were obtained on the two occasions.  The spectra shown in Figure~6
therefore represent the means of the epochs. HE~0236$-$2656 shows a deep,
broad ($\Delta\lambda\sim250$~\AA) feature centered near 5800~\AA, and there is
some evidence for the very shallow and extended depression around 4700~\AA\
that was noted by Reimers et al.  The circular polarisation is uniform at
$v\sim-1.5\%$ outside the $\lambda$5800 feature, but increases in magnitude to
nearly $-$5\% within the trough.  A single absorption line near $\lambda$6650
also dominates HE~0330$-$0002, but that feature is more triangular in
appearance and shows structure in the blue wing.  The circular polarisation in
HE~0330$-$0002 is predominantly negative over the spectral region observed and
becomes stronger toward the blue, but changes sign sharply to positive values
within the line. There may be an additional weak depression near 6100~\AA\ and
again around 4700~\AA. Together with a narrower line at $\sim$3900~\AA, the
latter appears as a more defined feature in the Reimers et al. spectrum.

The continuum energy distributions of both stars suggest that they are rather
cool, $T\sim$~$6000-7000$~K.  At these temperatures, He-atmosphere white
dwarfs are common, but spectral features are generally due to trace abundances
of atomic and/or molecular carbon and other metals.  The spectra of very few
of these species are known in magnetic fields of sufficient strength to
produce continuum circular polarisation of a few per cent ($B\gsimeq50$~MG).
For comparison, the reader is directed to LP~790-29 (8600~K; C$_2$ at 50~MG;
Liebert et al. 1978), G99-37 (6200~K; C$_2$, CH at $\sim$3~MG; Angel, Hintzen,
\& Landstreet 1975), LHS~2229 ($\sim$4600~K; C$_2$H? at $\sim$100~MG; Schmidt
et al. 1999), and LHS~2534 (6000~K; metallic atoms at 1.9~MG; Reid, Liebert,
\& Schmidt 2001). It seems likely that both HE~0236$-$2656 and HE~0330$-$0002
are He-dominated stars with yet other combinations of field strength and/or
atomic and molecular species.

\subsection{HE~0003$-$5701 \& HE~0338$-$3853: Non-magnetic DA + cool dwarf binaries}

Four stars were proposed as magnetic DB white dwarfs by Reimers et al. (1998)
on the basis of single low-resolution spectra.  All show flat to bluish
continua in $F_\lambda$, a rather narrow feature near $\lambda$5890 that was
assigned to the $\pi$ component of He~I $\lambda$5876, and more diffuse
absorption around $\lambda$5100 that was interpreted as He~I $\lambda$4921.
The implied field strength for each star was $\sim$20~MG.  Helium-atmosphere
stars comprise only $\sim$10\% of all white dwarfs, and at the time of these
claims, no magnetic white dwarf had been positively identified solely with He
features\footnote{As discussed in $\S$3.1, GD~229, HE~1211$-$1707 \&
HE~1043$-$0502 are now known to be magnetic DBs. Feige~7 (e.g. Achilleos et al.
1992), LB~11146 (Liebert et al. 1993) and LB8827 (Wesemael et al. 2001) are all
magnetic DBA (mixed hydrogen-helium) white dwarfs.}.  The report of 4 new
magnetic DBs was therefore remarkable.

Two of the proposed magnetic DB stars, HE~0003$-$5701 and HE~0338$-$3853, were
observed as part of this study.  Neither show circular polarisation at a level
exceeding $v=1\%$ anywhere in the spectrum, and the spectrum-added value for
HE~0003$-$5701 is an insignificant $-0.14$\% (Table~1).  Observations for
HE~0338$-$3853 were obtained under poorer observing conditions, so the results,
$\av{v}\sim\pm0.4\%$, should not be regarded as real detections. More
importantly, the spectra shown in Figures~7 and 8 reveal that the narrow
spectral line in both stars is actually Na~I D $\lambda\lambda$5890,5896 at
zero magnetic field, and the diffuse feature at shorter wavelengths is the
Mg~I/MgH complex prominent in late-type stars. From the detailed
correspondence between the host of absorption lines in the two Hamburg/ESO
objects and the cool dwarfs also shown in those figures, it is clear that both
proposed magnetic DB stars are actually binaries containing a late-type
component.  A third candidate, HE~0026$-$2150, was originally identified as a
binary on the basis of an \halpha\ emission line in the composite spectrum
(Reimers et al. 1998), but the true origin of the absorption features was not
inferred. Upon inspection of the original spectra, it now seems clear that
{\it all four\/} of the proposed magnetic DB stars -- HE~0003$-$5701,
HE~0026$-$2150, HE~0107$-$0158, and HE~0338$-$3853 -- are in fact non-magnetic
white dwarf + cool dwarf binary systems, and as such provide interesting
additions to lists of similar objects which have been assembled in recent
years (see, e.g., Marsh 2000; Vennes, Christian, \& Thorstensen 1998).

A dozen spectral templates of nearby Gliese dwarfs with spectral types K0 V $-$
M4 V were obtained for the purpose of decomposing the spectra of
HE~0003$-$5701 and HE~0338$-$3853, using the same instrumental setup and
during the same observing runs as for the magnetic white dwarf candidates. The
examples shown in Figures~7 and 8 represent our best attempts at identifying
the spectral type which quantitatively accounts for the observed features of
each star. The results of the subtraction are shown in the bottom panels of the
figures. The cleanest outcome is obtained for the fainter of the two objects,
HE~0338$-$3853, where the spectrum can be well-represented by a hot
(nonmagnetic) DA white dwarf and the K5($\pm$1) dwarf GL~186 shown in the
proportion 50:50($\pm$5) at 5500~\AA. Indeed, the only significant artifact in
the difference spectrum is a slight oversubtraction of the cool star for
$\lambda\gsimeq6900$~\AA. The inferred continuum shape and absorption lines
for the hot component can be compared to the DA star MCT 0455$-$281 (=EUVE
0457$-$281; also shown), which has been found by Vennes et al. 1997 to have a
temperature of 57,000~K. Based on \hbeta\ and \hgamma\ we cannot constrain the
surface gravity of the DA component in HE~0338$-$3853, but if we allow for
variations within the range $7.0\leq\log g\leq 9.5$, the effective temperature
is restricted to $87-97\times10^3$~K. The star could possibly be a hot sdO
similar to BD+28\gradi2411, but we find no evidence for He~II $\lambda$4686.

The hot star and its companion have comparable $V$-band magnitudes which we
estimate to be $V_{\rm K5V} = V_{\rm DA} \sim 17.7$.  The absolute magnitude
of a K5V star is $M_{\rm V, K5V} = 7.3$, so assuming that the two stars are
physically paired, the implied effective temperature of the hot star is
$\gsimeq80,000$~K and surface gravity $\log g \gsimeq 7.0-7.5$.  A lower
effective temperature would require a larger stellar radius, characteristic of
a hot sdO star.

The observed spectrum of HE~0003$-$5701 is redder and more highly-structured
than for HE~0338$-$3853, suggesting a larger fraction of the light is due to
the cool companion. Best cancellation of features occurs for a 60\%
contribution at $\lambda$5500 of the spectrum of GJ~3318.  This star is also
typed as K5~V but appears slightly redder than GL~186 in our data.
Uncertainties in the parameters of the decomposition are probably about twice
those for HE~0038$-$3853, i.e. $\sim$2 spectral subclasses and $\pm$0.1 in
relative brightness.  The difference spectrum shows the overall continuum shape
and smoothness of a hot white dwarf, but retains large-scale lumps suggestive
of a slightly incorrect template. Interestingly, \halpha\ emerges as a narrow
emission line regardless of the template used.  Emission cores may also be
present in \hbeta\ and \hgamma, though the latter is contaminated by Hg~I
$\lambda$4358 from Canberra city lights.  We cannot rule out intrinsic coronal
activity as the source of the Balmer emission, but EUV-illumination by the
white dwarf is a more likely explanation.  Radiative heating of the companion
may also modify its emergent energy distribution from that of an isolated dwarf
and contribute to difficulties with the spectral subtraction.  Similar
conclusions can be offered about the nature of the stellar components of this
binary as were made for HE~0338$-$3853 above, but the emission cores in the
white dwarf spectrum preclude a detailed line profile analysis.  Together with
HE~0026$-$2150, HE~0003$-$5701 presents a promising candidate for period
determination through radial velocity studies.  We note that the fourth
magnetic DB proposed by Reimers et al. (1998), HE~0107$-$0158, is so heavily
dominated by the hot component that, if it is also a white dwarf, the companion
must be quite cool.

\subsection{The unpolarised stars HE~0000$-$3430 \& HE~0127$-$3110}

The spectrum of HE~0000$-$3430 obtained by Reimers et al. (1996) is nearly
featureless longward of 5000~\AA, and the claim of a magnetic field ranging
between $\sim$40~MG and 120~MG over the stellar surface was based largely on
structure around 4700~\AA.  The star is comparatively bright, and our flux
spectra obtained on two consecutive nights each exhibit an extremely smooth
continuum, featureless to within fluxing irregularities, down to at least
4400~\AA\ (Figure~9). The circular polarisation is null in the
spectrum average to better than 0.15\% and there are no polarimetric features
or trends with wavelength to a per-pixel precision of $\sigma_v\lsimeq1\%$.
Since a magnetic white dwarf with a field strength near 100~MG and virtually
any viewing angle would be expected to show continuum polarisation
considerably larger than these limits, we cannot confirm the star as being
magnetic and can only suggest that the line identifications of Reimers et al.
(1996) may be spurious.  As those authors noted, the continuum slope of
HE~0000$-$3430 is rather shallow, indicating $T\sim7000$~K.  Assuming that the
star is indeed a white dwarf, the lack of spectral features suggests that it
is most likely one of the He-atmosphere examples that are common in this
temperature range.

HE~0127$-$3110 was observed on two occasions within a week.  Both the spectrum
and lack of circular polarisation are entirely consistent between the data
sets, so the combined results are shown in Figure~9.  We confirm the narrow
absorption feature around $\lambda$5890 reported by Reimers et al. (1996), but
find no evidence for the broad features around 4650~\AA\ and 5080~\AA\ that are
contained within our spectral range.  The circular polarisation shows no
spectral features and the coadded value is essentially zero at both epochs.
This is at odds with a magnetic field of $100-200$~MG as was required for the
model of Reimers et al.

The sole feature in our flux spectrum of HE~0127$-$3110 exhibits a red-shaded
profile extending from $\lambda$5850 to $\sim$$\lambda$5980, suggesting an
assignment with the Na~I D lines $\lambda\lambda5890,5896$ or possibly He~I
$\lambda$5876. The former are occasionally seen in the spectra of metallic-line
(DZ) white dwarfs. However, in G165-7 (Wehrse \& Liebert 1980), where the
6~\AA\ doublet splitting is not resolved due to pressure broadening, the
resulting feature is symmetric, quite unlike the profile smeared to the red in
Figure~9.  The observed profile cannot be attributed to magnetic effects, since
Na~I D is seen to form a classic symmetric Zeeman triplet with overall width
similar to that in Figure~9 in the magnetic white dwarf LHS~2534 ($B=1.9$~MG;
Reid et al. 2001). It is possible that the line profile in HE~0127$-$3110 is
contaminated by improper sky subtraction of the bright low-pressure Na~I D
lines from Canberra city lights, but the spectra of both nights show the same
distortion.  It should also be pointed out that both of the confirmed DZ stars
with Na~I absorption also show strong lines of Mg~I, Ca~I, and other metals.
If a Mg~I $\lambda5170$ complex is present in Figure~9, it is exceedingly
weak.  Finally, the stars where the Na~I lines are certain have $T_{\rm
eff}\sim~$$6000-7500$~K, significantly cooler than HE~0127$-$3110.  The He~I
$\lambda$5876 explanation faces problems as well, since features of
$\lambda$4472 and $\lambda$6678 of would be expected at similar strength for a
wide range of temperature (Wegner \& Nelan 1987). Thus, even though the
absence of polarisation suggests that the star is nonmagnetic, line
identification remains uncertain.

\section{Summary and conclusions}

Spectropolarimetric observations of several suspected southern magnetic white
dwarfs have confirmed the existence of substantial fields on HE~1211$-$1707,
HE~1043$-$0502, HE~1045$-$0908, HE~0236$-$2656 and HE~0330$-$0002.  The new
data also clearly show that the proposed magnetic DB white dwarfs
HE~0003$-$5701, HE~0026$-$2150, HE~0107$-$0158, and HE~0338$-$3853 are actually
nonmagnetic DA plus cool dwarf pairs.  Finally, two stars, HE~0000$-$3430 and
HE~0127$-$3110, are found to be polarimetrically null, but the latter remains
interesting due to the presence of an unidentified broad absorption line in its
spectrum.

Perhaps the most important result is the addition of HE~1211$-$1707
($\sim$50MG) and HE~1043$-$0502 ($\sim$800~MG) to the class of magnetic
DB stars along with the prototype of the class, GD~229.  Though He~I features
have long been assumed to be present in certain existing magnetic white dwarfs,
only now is it possible to make firm identifications of the highly shifted
spectral features, thanks to the development of computational techniques for
solving the multi-electron problem in magnetic fields of arbitrary strength
(e.g., Becken \& Schmelcher 1998).  Including LB~8827 (Wesemael et al. 2001),
Feige 7 (Achilleos et al. 1992), and GD~229 (Jordan et al. 1998), the spectrum
of neutral helium has now been empirically verified over the full range $1
\lsimeq B \lsimeq 800$~MG, essentially the same as is spanned by magnetic DAs.
The number of white dwarfs that evidence magnetic He~I lines amounts to
$\sim$10\% of the total magnetic sample, roughly consistent with the fraction
of DB+DBAs among white dwarfs in general, if the small number statistics and
probable selection effects are considered.

Our detailed understanding of the field structures on magnetic white dwarfs
will continue to improve with the evolution of more sophisticated modeling
techniques for the often non-dipolar field geometries which are now known to
exist over many stars.  The recovery of this information requires
high-quality, full phase-coverage, spectroscopic and spectropolarimetric data,
and in the case of the helium-line stars, the calculation of some wavelengths
as well as oscillator strengths for all transitions over a wide range in
magnetic field.  The results presented here should help to fuel the interest in
remedying both of these limitations.

\section{Acknowledgements}

G.S. is grateful for the hospitality and stimulating environments of the
Australian National University and Mt. Stromlo Observatory, where this work
was carried out during a sabbatical leave.  The assistance of the Mt. Stromlo
Instrument Shops in adapting the spectropolarimeter to the telescope is
especially appreciated.  Thanks also go to R. Deacon and A. Kawka for
assistance at the telescope. Studies of strongly magnetic stars and stellar
systems is supported at Steward Observatory by the National Science Foundation
through grant AST 97-30792.  S. Vennes is a QEII fellow of the Australian
Research Council.

\section*{References}

\reference Achilleos, N., Wickramasinghe, D.T., Liebert, J., Saffer, R.,
Grauer, A.D. 1992, ApJ, 396, 273

\reference Angel, J.R.P., Hintzen, P., Landstreet, J.D. 1975, ApJ, 196, L27

\reference Becken, W., Schmelcher, P., Diakonos, J. 1999, J. Phys. B., 32, 1557

\reference Becken, W., Schmelcher, P. 1998, in Atoms and Molecules in Strong
External Fields, ed. P. Schmelcher and W. Schweizer (New York: Plenum), 207

\reference \sameauthor\ 2000, J. Phys. B., 33, 545

\reference \sameauthor\ 2001, Phys. Rev. A, 63, 053412

\reference Hakala, P.J. 1995, A\&A, 296, 164

\reference Jordan, S. 2001, in 12th European Workshop on White Dwarfs,
ASP Conf. Ser. Vol. 226, ed. J.L. Provencal, H.L. Shipman, J. MacDonald, \&
S. Goodchild (San Francisco: ASP), 269

\reference Jordan, S., Schmelcher, P., Becken, W., Schweizer, W. 1998, A\&A,
336, L33

\reference Kemic, S.B. 1974, JILA Rep., 113

\reference Liebert, J., Bergeron, P., Schmidt, G.D., Saffer, R.A. 1993, ApJ,
418, 426

\reference Liebert, J., Angel, J.R.P., Stockman, H.S., Beaver, E.A. 1978, ApJ,
225, 181

\reference Marsh, T.R. 2000, New Astronomy Reviews, 44, 119

\reference  Maxted, P.F.L., Ferrario, L., Marsh, T.R., Wickramasinghe, D.T.
2000, MNRAS, 315, L41

\reference Reid, I.N., Liebert, J., Schmidt, G.D. 2001, ApJ, in press

\reference Reimers, D., Jordan, S., Beckmann, V., Christlieb, N., Wisotzki
1998, A\&A, 337, L13

\reference Reimers, D., Jordan, S., Koester, D., Bade, N., K\"ohler, Th.,
Wisotzki, L. 196, A\&A, 311, 572

\reference Reimers, D., Jordan, S., K\"ohler, T., Wisotzki, L. 1994, A\&A, 285 995

\reference Schmidt, G.D., West, S.C., Liebert, J., Green, R.F., Stockman,
H.S. 1986, ApJ, 309, 218

\reference Schmidt, G.D., Liebert, J., Harris, H.C., Dahn, C.C., Leggett, S.K.
1999, ApJ, 512, 916

\reference Schmidt, G.D., Stockman, H.S., Smith, P.S. 1992, ApJ, 398, L57

\reference
\reference Vennes, S., Christian, D.J., Thorstensen, J.R. 1998, ApJ, 502, 763

\reference Vennes, S., Thejll, P., Galvan, R.G., Dupuis, J. 1997, ApJ, 480, 714

\reference Wegner, G., Nelan, E.P. 1987, ApJ, 319, 916

\reference Wehrse, R., Liebert, J. 1980, A\&A, 86, 139

\reference Wesemael, F., Liebert, J., Schmidt, G.D., Beauchamp, A., Bergeron,
P., Fontaine, G. 2001, ApJ, in press

\reference Wickramasinghe, D.T., Ferrario, L. 2000, PASP, 112, 873

\reference{} Wickramasinghe, D.T., Martin, B. 1979, MNRAS, 188, 165

\reference Wisotzki, L., K\"ohler, T., Grote, D., Reimers, D., 1995, A\&AS,
115, 227


\begin{table*}
 \centering
 \begin{minipage}{140mm}
  \caption{Log of Observations.}
  \begin{tabular}{@{}cccccl@{}}
Object     & $m_{5500}$ & UT Date & \#Obs$\times$Expo(s)\footnote{Number of waveplate sequences
and exposure time per sequence} & $\av{v}(\%)$\footnote{Spectrum-summed circular polarisation}
& Comment \\[10pt]
HE 0000$-$3430 & 15.0 & 2000 Dec. 31 & 2\by720   &   +0.05 & cirrus \\
               &      & 2001 Jan. ~\,1 & 2\by720 & $-$0.14 & \vspace{6pt} \\
HE 0003$-$5701 & 15.5 & 2000 Dec. 28 & 2\by720   &   +0.14 & \vspace{6pt} \\
HE 0127$-$3110 & 16.1 & 2001 Feb. 18 & 2\by720   & $-$0.09 & \\
               &      & 2001 Feb. 22 & 3\by720   & $-$0.07 & \vspace{6pt} \\
HE 0236$-$2656 & 17.1 & 2001 Feb. 17 & 3\by720   & $-$1.27 & \\
               &      & 2001 Feb. 20 & 2\by720   & $-$1.19 & \vspace{6pt} \\
HE 0330$-$0002 & 16.9 & 2001 Feb. 18 & 2\by720   & $-$1.42 & \\
               &      & 2001 Feb. 23 & 2\by720   & $-$0.90 & cirrus \vspace{6pt} \\
HE 0338$-$3853 & 17.0 & 2000 Nov. 25 & 4\by800   & $-$0.41 & cirrus \\
               &      & 2000 Nov. 26 & 2\by800   &   +0.44 & poor seeing \vspace{6pt} \\
HE 1043$-$0502 & 17.1 & 2001 Feb. 18 & 9\by720   &   +1.49 & \\
               &      & 2001 Feb. 22 & 9\by720   &   +1.62 & \vspace{6pt} \\
HE 1045$-$0908 & 16.4 & 2001 Feb. 21 & 5\by720   & $-$0.07 to $-$0.87 & rotating $P\sim2-4$~hr \vspace{6pt} \\
HE 1211$-$1707 & 17.1 & 2001 Feb. 22 & 15\by360  & +0.94   & rotating $P\sim2$~hr \\
               &      & 2001 Feb. 25 & 15\by360  & +1.22   &
\end{tabular}
\end{minipage}
\end{table*}

\clearpage

\begin{figure}
\includegraphics{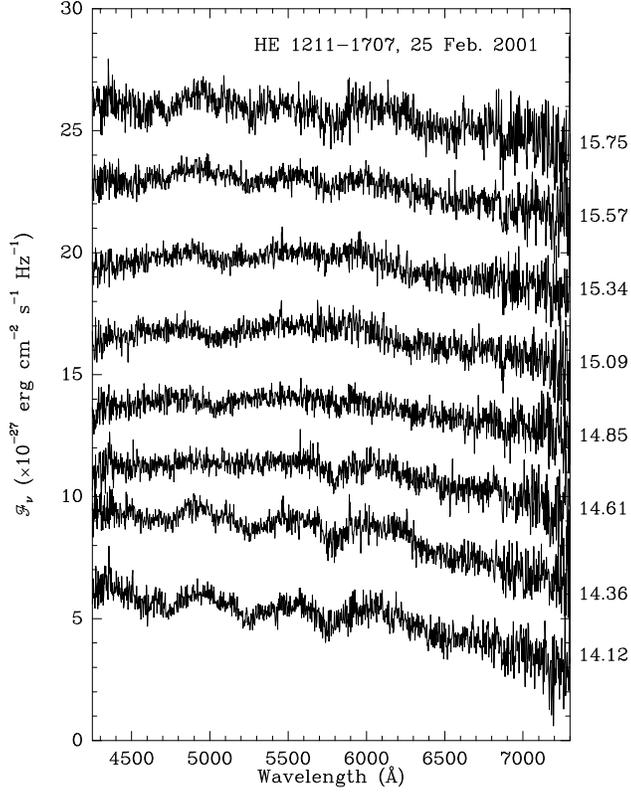}
\vspace{25pc}

\caption{Spectral sequence for HE~1211$-$1707 obtained on 25 Feb. 2001.  UT is noted
at the right.}

\label{}
\end{figure}

\begin{figure}
\includegraphics{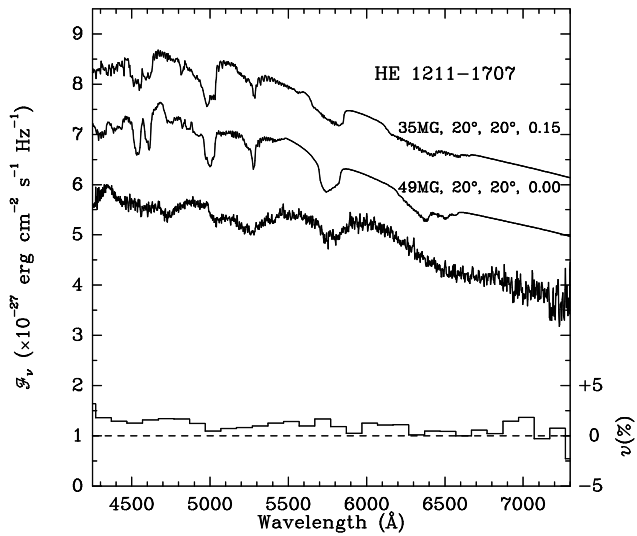}
\vspace{18pc}

\caption{Mean spectrum of HE~1211$-$1707, averaged over both nights and shown
as $F_\nu$ to enhance the weak spectral features.  For comparison are plotted
two crude He-atmosphere spectral models, labeled by the dipolar magnetic field
strength, inclination, azimuth, and longitudinal field offset (in units of the
stellar radius), respectively.  At the bottom is the phase-averaged observed
circular polarisation, binned into 100~\AA\ increments.}

\label{}
\end{figure}

\begin{figure}
\includegraphics{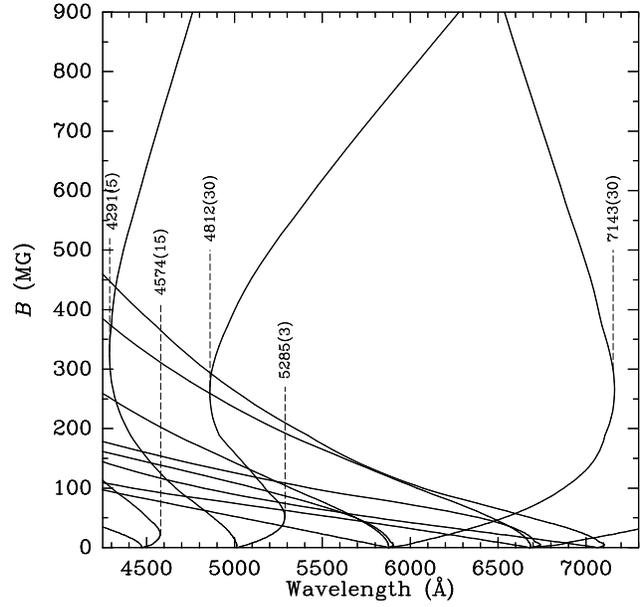}
\vspace{20pc}

\caption{Dependence of major transitions of He~I on magnetic field strength for
$0<B<900$~MG, from the calculations of Becken \& Schmelcher (1998, 2000, 2001)
and Becken, Schmelcher, \& Diakonos (1999).  Wavelengths and uncertainties are
provided for turnaround points, where the strongest spectral features are to be
expected.}

\label{}
\end{figure}

\begin{figure}
\includegraphics{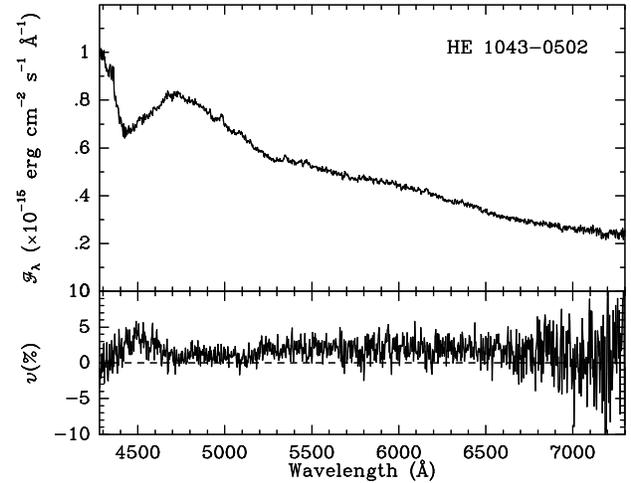}
\vspace{17pc}

\caption{Total flux {\it (top)\/} and coadded circular polarisation {\it
(bottom)\/} spectra of HE~1043$-$0502.  Comparison of the observed spectrum
with the $B(\lambda$) curves in Figure~3 suggest that the features are He~I
lines in a magnetic field $B\sim450-800$~MG.}

\label{}
\end{figure}

\begin{figure}
\includegraphics{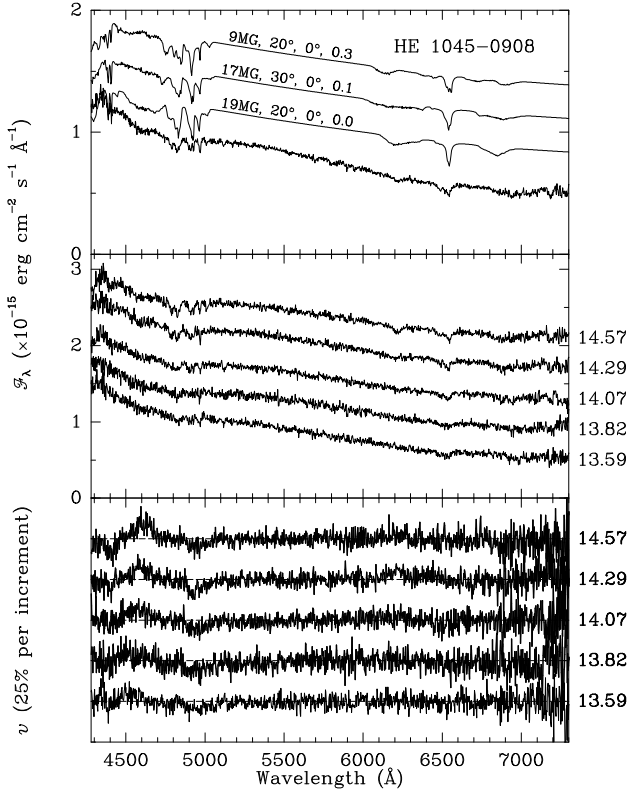}
\vspace{26pc}

\caption{Sequences of total flux {\it (middle)\/} and circular polarisation
{\it (bottom)\/} for HE~1045$-$0908, obtained 2001 Feb. 21.  A rotation period
of a few hours is indicated by the variations with UT (hrs), noted at the
right. {\it Top panel:} Observed phase-averaged spectrum and 3 model spectra
for the indicated dipolar magnetic field strength, inclination, azimuth, and
longitudinal field offset.}

\label{}
\end{figure}

\begin{figure}
\includegraphics{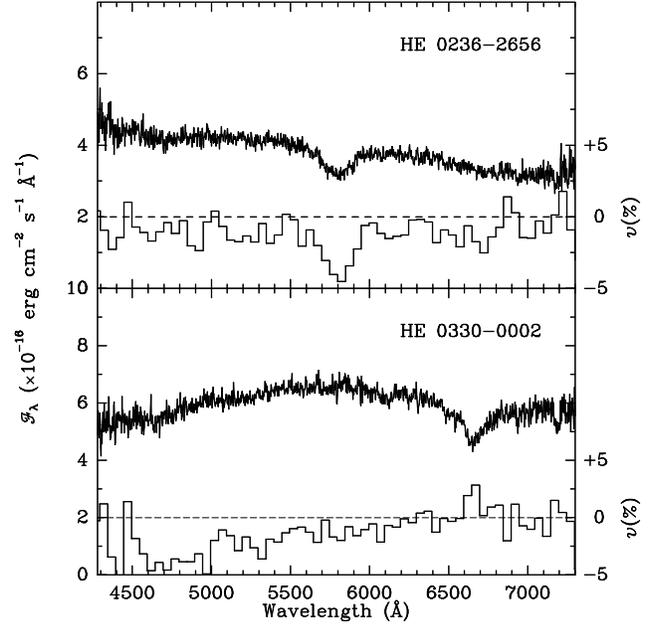}
\vspace{21pc}

\caption{{\it (Top):\/} Flux and circular polarisation spectra for
HE~0236$-$2656, the latter binned into 50~\AA\ increments.  The polarisation is
small and uniformly negative outside the $\lambda$5800 absorption feature, but
increases to $v\sim-5\%$ within the trough. {\it (Bottom):\/} As in the top
panel for HE~0330$-$0002.  The polarisation becomes increasingly negative with
decreasing wavelength and changes sign within the broad $\lambda$6650
absorption feature. Subtle depressions in spectral flux may also exist around
$\lambda$4700 and $\lambda$6100. }

\label{}
\end{figure}

\begin{figure}
\includegraphics{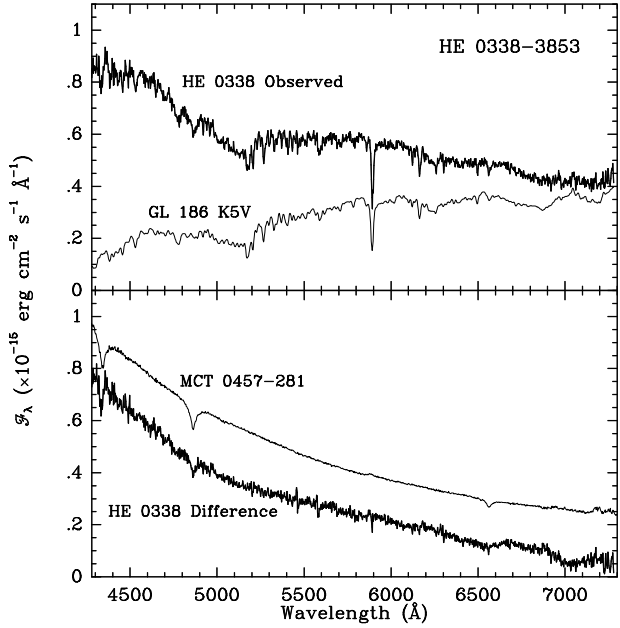}
\vspace{21pc}

\caption{{\it (Top):\/} Observed total flux spectrum of HE~0338$-$3853
compared with the K5~V dwarf GL~186.  The detailed correspondence of spectral
features throughout the spectra attest to the presence of a late-type
component. {\it (Bottom):\/} Difference spectrum portraying the white dwarf
component. A representative hot DA white dwarf, MCT 0457$-$281, is also
shown, displaced upward for clarity.}

\label{}
\end{figure}

\begin{figure}
\includegraphics{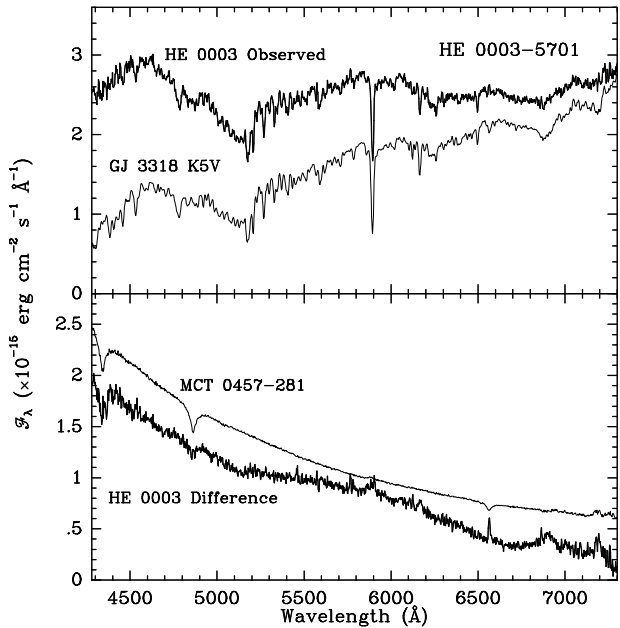}
\vspace{21pc}

\caption{As in Figure~7 for HE~0003$-$5701.  Narrow emission at \halpha\
signifies illumination of the cool dwarf in a close binary system.}

\label{}

\end{figure}

\begin{figure}
\includegraphics{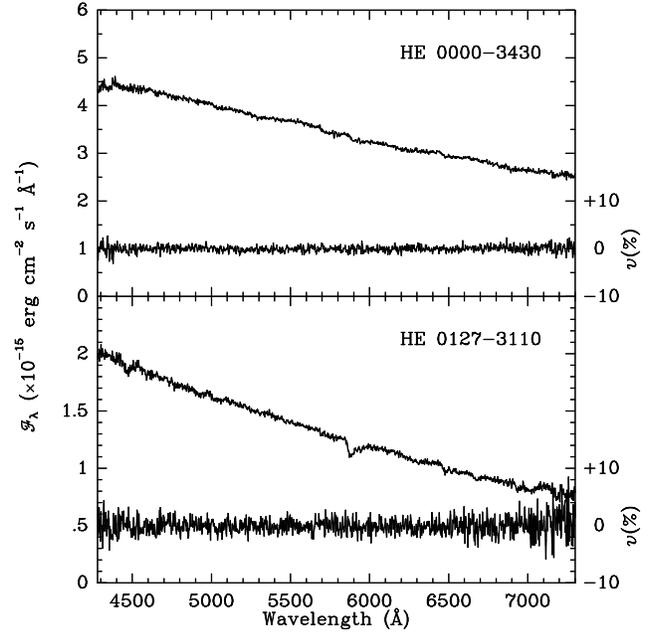}
\vspace{21pc}

\caption{{\it (Top):\/} HE~0000$-$3430.  Underlying the featureless (DC) flux
spectrum is the circular polarisation spectrum, with units indicated on the
scale at the right. {\it (Bottom):\/} A single feature is detected near
5890~\AA\ in HE~0127$-$3110. Like HE~0000$-$3430, the object shows no
significant circular polarisation.}

\label{}

\end{figure}

\end{document}